\documentclass[runningheads]{llncs}
\usepackage{soul}
\usepackage{color}

\usepackage[T1]{fontenc}
\usepackage{graphicx}
\usepackage{booktabs}
\usepackage{comment}
\usepackage{tabularx}
 \usepackage{array}
 \newcolumntype{Y}{>{\raggedright\arraybackslash}X}

\usepackage{ifthen}

\newboolean{auxinfo}
\setboolean{auxinfo}{true}   

\newboolean{showoutline}
\setboolean{showoutline}{true}

\newboolean{showcomment}

\setboolean{showcomment}{true} 

\usepackage[moderate]{savetrees}

\ifthenelse{\boolean{auxinfo}}
  {}
  {
    \setboolean{showoutline}{false}
    \setboolean{showcomment}{false}
  }

\usepackage{xcolor}
\usepackage{amssymb}

\ifthenelse{\boolean{showcomment}}
   {\newcommand{\nb}[2]{\fcolorbox{gray}{yellow}{\bfseries\sffamily\scriptsize#1}{\sf\small$\blacktriangleright${\em #2}$\blacktriangleleft$}}
   \newcommand{\working}[1]{\fcolorbox{gray}{yellow}{{\bf #1}\emph{\scriptsize---in progress---}}}
   \newcommand{\TBD}[1]{\fcolorbox{gray}{yellow}{{\bf #1}\textbf{TBD}}} 
  }
  {\newcommand{\nb}[2]{}{}
   \newcommand{\working}[1]{}
   \newcommand{\TBD}[1]{} 
  }

\usepackage{todonotes}
\ifthenelse{\boolean{showoutline}}{
	\newcommand{\outline}[3]{
		~\newline 
		\fcolorbox{red}{white}{
			\parbox{\columnwidth}{
				\ifthenelse{\equal{#1}{}}{
					\ifthenelse{\equal{#2}{}}{
						\noindent\colorbox[rgb]{0.65,0.16,0}{\textcolor[rgb]{1,1,1}{\textbf{Outline}}}
					}{
						\colorbox[rgb]{0.65,0.16,0}{\textcolor[rgb]{1,1,1}{\textbf{Outline -- Responsible: #2}}}
					}
				}{
					\ifthenelse{\equal{#2}{}}{
						\noindent\colorbox[rgb]{0.65,0.16,0}{\textcolor[rgb]{1,1,1}{\textbf{#1 page(s)}}}
					}{
						\colorbox[rgb]{0.65,0.16,0}{\textcolor[rgb]{1,1,1}{\textbf{#1 page(s) -- Responsible: #2}}}
					}
				}
				#3
			}
		}
	}
}{
	\newcommand{\outline}[3]{}
}

\newcommand\defauxcomm[1]{
       \expandafter\newcommand\csname #1\endcsname[1]{\nb{#1}{##1}}
       \expandafter\newcommand\csname WK#1\endcsname{\working{#1}}
       \expandafter\newcommand\csname TBD#1\endcsname{\nb{#1}}
    } 
   
\defauxcomm{KS}
\defauxcomm{MN}

\usepackage[normalem]{ulem}
\ifthenelse{\boolean{showcomment}}
    {\newcommand{\strike}[1]{\textcolor{red}{\sout{#1}}}}
    {\newcommand{\strike}[1]{}}

\begin{document}

%
\title{Beyond Establishing the Four-Day Workweek:\\ Understanding Adaptation and Long-Term Survival in an Agile Software Organization}
\titlerunning{Beyond Establishing the Four-Day Workweek}
%
\author{Michael Neumann\inst{1}\orcidID{0000-0002-4220-9641}\and
Darja \v{S}mite\inst{2}\orcidID{0000-0003-1744-3118}
}
\authorrunning{M. Neumann and D. \v{S}mite}
%
\institute{University of Applied Sciences \& Arts Hannover,\\ Hannover, Germany 
\email{michael.neumann@hs-hannover.de}\\
\and
Blekinge Institute of Technology,  Karlskrona , Sweden
\email{darja.smite@bth.se}
}

\maketitle              
\begin{abstract}
\textit{Context:} Existing research on the four-day workweek (4DWW) has primarily examined its introduction and short-term effects, with limited understanding of its long-term survival or its interaction with agile software development.
\textit{Objective:} We study how a reduced-hour 4DWW is introduced, adapted, institutionalized, and sustained under changing organizational and external conditions in an agile software organization.
\textit{Method:} We conducted a longitudinal single-case study of a software organization operating a 32-hour, four-day week. The study draws on 15 semi-structured interviews in 2022 and 2026, analyzed using qualitative content analysis. 
\textit{Results:} The 4DWW is better understood as an evolving arrangement than a one-off intervention. After the introduction, teams redesigned  coordination, communication, meetings, agile practices, and iterations to adapt to reduced working time. Once institutionalized, the 4DWW faced ownership change, economic downturn, and market and AI pressures. Rather than reverting to five-day workweek, employees absorbed these pressures through voluntary protective adaptations, while anticipating that a rollback would harm job satisfaction, organizational commitment, and employer image.
\textit{Contribution:} We contribute longitudinal evidence on sustaining 4DWW in agile software development and propose two conceptual artifacts: a lifecycle model of the 4DWW and a 4DWW survival matrix, explaining how external pressures and management rationale rollback risk.

\end{abstract}
\section{Introduction}
The four-day workweek (4DWW) has moved from an experimental concept to a working-time model. 4DWW is debated and adopted across various industries over the past decades~\cite{Veal.2023}. National pilots\footnote{An overview of conducted pilots of the 4DWW is given by Ahmed et al.~\cite{Ahmed.2025}.} in Iceland~\cite{Joyeaux.2025}, Belgium~\cite{Dutordoir.2024}, and the United Kingdom~\cite{Lewis.2023}, alongside private-sector adoptions~\cite{Oliveira.2025}, have placed reduced working time on the agenda of politicians, practitioners, and researchers~\cite{Campbell.2023,Landwehr.2025}. Even GenAI producers, like OpenAI, motivate employers to convert efficiency dividends gained from AI adoption into investments in workers' benefits, with 32-hour/four-day workweek being one important solution to consider\footnote{OpenAI's Industrial Policy for the Inteligence Age: Ideas to Keep People First \cite{OpenAI.2026}.}.

The interest in 4DWW is motivated by the positive effects of 4DWW on productivity, well-being, retention, and organizational performance across industries, as cataloged in systematic reviews on the topic~\cite{Jahal.2024}. Empirical studies report increased focus on tasks with high-value outcomes and reductions in meeting overhead~\cite{Chakraborty.2007,Munyon.2023}. On the practice side, the drivers for 4DWW include talent competition in tight labor markets, post-pandemic re-evaluations of working-time norms, and the normalization of distributed and hybrid work in software organizations~\cite{Smite.2023}. 

Software-intensive organizations, known for their interest in innovative work models and employee engagement, are a particularly relevant context for studying the 4DWW. At the same time, the path toward a 4DWW in companies following agile practices isn't straightforward. Software engineering organizations are known for being dynamic work environments marked by tight deadlines, high workload, task interdependence, and uncertainty in general~\cite{godliauskas.2025}, and especially in agile contexts ~\cite{venkatesh2020agile}. Agile software development further relies on recurring collaborative practices, such as Daily Standups, Iteration Reviews, and Retrospectives, that structure coordination, feedback, and decision-making in time-boxed cycles~\footnote{In this paper, we follow the terminology based on the taxonomy of agile elements~\cite{Neumann.2021} and understand an agile practice as an applied activity by using tools and methods.}. Reducing the work time into four days can therefore affect not only working time, but also team coordination, synchronization, decision-making, and product delivery cycles. Agile software organizations thus appear simultaneously well-suited to and potentially vulnerable under a 4DWW.

Despite the growing interest in the 4DWW, existing research has primarily focused on the immediate outcomes following its introduction~\cite{Jahal.2024}. While these studies provide valuable insights into the effects of reduced working time, they often portray the 4DWW as a stable organizational state. In practice, however, organizations continue to evolve. Leadership changes, market pressure, economic downturns, workforce turnover, and shifting customer demands may all challenge the viability of an established 4DWW. Thus, the question is no longer only whether organizations should or can successfully introduce a 4DWW, but also how such arrangements adapt, institutionalize, and survive over time.

This question is particularly important for 4DWW models that reduce rather than compress working time.  Unlike compressed workweek arrangements that redistribute the same number of hours across fewer days, the case company studied in this paper adopted a 32-hour workweek over four days, with employees typically working 7.5 to 8.5 hours per day. Such an arrangement requires the organization to sustain customer responsiveness, coordination, and product delivery objectives with reduced available working time. While previous studies have examined experiences with the 4DWW in agile software development contexts~\cite{Topp.2022}, little is known about how such arrangements are sustained over multiple years, how organizations respond to emerging pressures, and how employees perceive the possibility of returning to a five-day workweek (5DWW). 

To address this gap, we conducted a longitudinal case study of an agile software organization that has operated under a 4DWW for several years. Rather than focusing solely on the transition itself, we examine the broader lifecycle of the 4DWW, including its introduction, adaptation, institutionalization, and response to organizational and external pressures. Through this perspective, we seek to answer the following two research questions:

    \textit{\textbf{RQ1: How is a four-day workweek introduced, adapted, and institutionalized in an agile software organization?}}\\
    Our first RQ examines the organizational and behavioral processes through which a 4DWW normalizes in an organizational context. It focuses on the conditions supporting adoption, the adaptations required in everyday work practices, and the mechanisms through which the 4DWW becomes embedded in organizational routines and employee expectations.

    \textit{\textbf{RQ2: How does a four-day workweek survive organizational and external pressures, and what factors may lead to its rollback?}}\\
    Our second RQ examines how organizations and employees respond when an established 4DWW comes under pressure from changing business conditions, market demands, or organizational restructuring. It focuses on resilience mechanisms, protective adaptations, and the anticipated consequences of reverting to a 5DWW.

The rest of the paper is structured as follows: In Section~\ref{sec:RelWork}, we give an overview of the related work to identify our research gap. Next, we explain our research study design in Section~\ref{sec:ResearchDesign}, including a detailed description of the case context and their applied 4-day work-week approach. We answer our research questions in Section~\ref{sec:Results} and discuss our results in Section~\ref{sec:Discussion}. The paper closes with a conclusion and an outlook on our future work activities in Section~\ref{sec:Conclusion}.

\section{Background \& Related Work}
\label{sec:RelWork}
In this section, we first explain the fundamentals of the 4DWW concept emphasizing its different application types, 
followed by an overview of scientific literature, in particular primary and secondary peer-reviewed studies, dealing with 4DWW phenomenon. 
\vspace{-4mm}
\subsection{The 4DWW Concept and Its Introduction}
The 4DWW is not a recent phenomenon. Discussions of shortened work weeks appear in labor research as early as the 1950s, and the first organizational experiments at scale were conducted in the United States during the 1970s~\cite{Hartman.1977}. European labor markets followed a different trajectory in the same period, emphasizing flexible work hours and other forms of working-time flexibilization rather than reduced weekly days~\cite{Campbell.2023,Veal.2023}. Sustained international attention to the 4DWW intensified only in the past decade, driven by post-pandemic reassessments of attitudes toward working time and by changing workforce expectations regarding well-being and work-life integration~\cite{Landwehr.2025,Veal.2023}.

The literature does not describe a single or standardized 4DWW model. In their systematic review, Landwehr et al.~\cite{Landwehr.2025} catalog several conceptualizations that differ in whether weekly working hours are genuinely reduced or only compressed into four longer days, whether compensation remains constant, and whether the free day is fixed or rotating. The most visible variant in current public debate combines unchanged pay with reduced working time and an expected productivity level equal to the previous five-day week, often labeled 100-80-100. Other variants retain the original 40-hour week through compressed 4-by-10 schedules or rely on staggered patterns that keep operations covered across all weekdays~\cite{Landwehr.2025}. Reported effects on productivity, well-being, and retention vary depending on the specific model adopted~\cite{Campbell.2023,Landwehr.2025}, so the choice of variant precedes the question of effects.

These design choices also shape what an introduction requires of an organization. Cruz et al.~\cite{Cruz.2026} report on multiple organizations and observe that formal 4DWW policy and lived team practice diverge along lines of organizational culture. Tensions surface between managers, employees, and operational requirements once the schedule change moves from announcement to daily work, and the introduction therefore extends well beyond the formal transition date. Munyon et al.~\cite{Munyon.2023} identify managerial support and redesigned workflows as conditions associated with a successful introduction of compressed-week models. A compressed week retains weekly working hours and therefore differs from a reduced-hour 4DWW, but the underlying argument that organizations need to adjust work routines before rather than after the schedule change recurs in broader 4DWW accounts.
\vspace{-4mm}
\subsection{The 4DWW in Software Engineering Organizations}
Once a 4DWW is in place, the relevant question shifts from initial adoption to how the model interacts with everyday work. Evidence on this post-introduction phase, and on the agile software engineering context in particular, is thin.

Topp et al.~\cite{Topp.2022} report on an agile software development organization in the period following a 4DWW introduction. They document changes in remote work patterns, team coordination, and individual perceptions of productivity, but their account is cross-sectional and limited to a single time point shortly after the transition. Neumann et al.~\cite{Neumann.2025} study an established compressed work schedule in a software engineering setting and report perceived gains in concentration alongside developer-relevant stress signals. Their case differs from a reduced-hour 4DWW in that weekly hours remain constant, so the workload-per-day implication does not transfer directly.

Across sectors, evidence remains largely short-term. Jahal et al.~\cite{Jahal.2024} note that most studies in their review measure outcomes only months after the transition, documenting immediate effects but rarely assessing their long-term sustainability under shifts in workforce composition, market pressure, or leadership.

Within software engineering, two aspects of the post-introduction phase have received little attention so far. One concerns how recurring synchronous practices such as Daily Standups, Review Meetings, and Retrospectives are adapted once the fifth working day is removed for an extended period. The other concerns which organizational conditions sustain the 4DWW over time. Our longitudinal single-case study contributes evidence on both aspects through data collected at two points in time more than three years apart within the same organization.

\section{Research Design}
\label{sec:ResearchDesign}
\subsection{Single Case Study \& Context Description}
In this paper, we report our results from a longitudinal single-case study~\cite{Yin.2018} examining an organizational transition to a 4DWW at the pseudonymous company Pied Piper. The primary unit of analysis is the lifecycle of the four-day workweek. To understand this lifecycle, we examined the organizational conditions, work practices, employee responses, and external pressures that shaped how the 4DWW was introduced, adapted, normalized, and sustained over time. The study follows an embedded case study design~\cite{Yin.2018}, where the overall case is the implementation and evolution of the 4DWW, while individual employees, teams, and organizational functions serve as embedded units of analysis through which the broader organizational dynamics are examined.

Pied Piper is operating worldwide in the online marketing industry and employs around 1,450 employees. The software development sites are located in Germany, Poland, Romania, and the United Kingdom. Our study focuses on one of the software development departments in Germany and its interface roles in Italy.

\subsection{Data Collection \& Analysis}
For this study, we applied a qualitative data collection and analysis approach according to the guideline by Yin~\cite{Yin.2018}. The data collection was divided into two phases.

The first phase consisted of 10 semi-structured interviews conducted in June, November, and December 2022 with the employees of Pied Piper covering all agile roles and various levels of experiences working for Pier Piper. Table~\ref{tab:interview_participants} provides detailed information of the interviewees per phase. For the first phase, we used an interview guideline, which we created by using the Goal Question Metric (GQM)~\cite{Basili.1994}. The guideline consisted of 34 questions: six questions regarding the interviewee background, 17 dealing with the perceived effects of the 4DWW transformation on individual and team level, five on a possible backward compatibility, and four closing questions. We made the interview guideline available in the research protocol~\cite{Neumann.2026}. In order to ensure the quality of the interview guide, we conducted a pilot interview (not recorded) to identify possible optimization measures, however, none were needed, so we decided to include the interview with interviewee P01 into the analysis process. The interviews of the first phase lasted 20-81 minutes (average of 45 minutes).  

\begin{table}[htbp]
\centering
\caption{Overview of interview participants}
\label{tab:interview_participants}
\begin{tabular}
{p{1cm} p{1cm} p{4.2cm} p{2cm} p{1.8cm} p{1.8cm}}
\toprule
\textbf{Phase} & \textbf{ID} & \textbf{Role} & \textbf{Date} & \textbf{Duration (min)}  & \textbf{Language} \\
\midrule
1  & ID1  & Senior Agile Coach         & 07.06.2022 & 58:21 &  German\\
1  & ID2  & Lead Engineer Dev Team    & 10.11.2022 & 55:32 &  German\\
1  & ID3  & IT Project Manager         & 14.11.2022 & 81:07 &  German\\
1  & ID4  & Department Lead Hannover   & 23.11.2022 & 26:07 &  German\\
1  & ID5  & Software Developer         & 24.11.2022 & 22:12 &  German\\
1  & ID6  & Senior Quality Manager     & 30.11.2022 & 38:30 &  German\\
1 & ID7 & Product Manager           & 30.11.2022 & 52:40 &  English\\
1  & ID8  & Software Architect         & 01.12.2022 & 43:02 &  German\\
1  & ID9  & Software Developer        & 05.12.2022 & 43:54 &  German\\
1  & ID10  & Software Developer        & 12.12.2022 & 30:33 &  German\\
2 & ID11 & Lead Engineer Dev Team    & 13.05.2026 & 38:30 &  English\\
2 & ID12 & Software Engineer          & 18.05.2026 & 38:06 &  German\\
2 & ID13 & IT Project Manager         & 19.05.2026 & 25:27 &  German\\
2 & ID14 & Product Manager            & 21.05.2026 & 34:32 &  English\\
2 & ID15 & Head of Agile Coaches      & 26.05.2026 & 28:08 &  German \\
\bottomrule
\end{tabular}
\end{table}

The second phase comprised five semi-structured interviews conducted via MS Teams in May 2026. Interviews lasted 25-38 minutes (average of 33 minutes). As the second phase was designed covering the longitudinal effects of the 4DWW, we decided to adapt the existing interview-guide. To be more precise, we shortened the guideline by asking demographic information updates in the past years, especially related to career aspects and focusing strongly on the observed and perceived changes related to the 4DWW by the participants. The adapted interview guide consists of 18 questions and can be found in the research protocol~\cite{Neumann.2026}. 

All interviews except P01 were held virtually via Microsoft (MS) Teams, recorded and automatically transcribed using MS Teams functionality. Transcripts in German language were translated into English. The quality of the transcripts was ensured by validating the transcript shortly after conducting the interview to ensure completeness and verifying no hallucinated content was generated. 

We analyzed the data using qualitative content analysis according to Mayring~\cite{Mayring.2014}. In the first step, we read the transcripts to familiarize with the data, noted and discussed primary codes and preliminary categories (e.g., leadership support, economic context). Finally, we clustered these categories under the themes based on the research questions (e.g., Introduction of 4DWW). An overview of the identified themes, categories, and the count of mapped codes across the interview phases is given in Table~\ref{tab:coding}.

Coding across the two interview phases varied. Analysis of the first phase interviews was done by the first author manually. For the second phase, both authors coded disjoint subsets independently and double-coded an overlap of two interviews. Comparison of the independent coding results showed significant overlap of emerging categories. We additionally coded all interviews using Artificial Intelligence (AI) tools, in particular Opus 4.8 and Fable 5 Large Language Models (LLMs) using Claude Desktop App as well as ChatGPT GPT 5.5. The LLMs were applied as a secondary annotator for quality assurance of the coding in accordance to Baltes et al. \cite{Baltes.2025,Baltes.2026}. We treated the model output as a complementary consistency check rather than an independent reliability measure. Based on these analysis results, we compared the codes, categories, and themes generated by the LLMs with our own results and observed no substantive differences in the resulting category system. For both the codebook and the quotes reported in this paper, we relied on our manual coding, as it retained substantially more surrounding context than the passages identified by the LLM. The model-assisted analysis was run under data-processing configuration that excludes the interview data from model training. For data analysis and documentation, we used MS Word and Excel. The codebook is available in the research protocol~\cite{Neumann.2026}.

\begin{table*}[htbp]
\centering
\small
\caption{Overview of the qualitative coding structure with frequencies across the two interview phases (P1 and P2)}
\label{tab:coding}
\begin{tabularx}{\textwidth}{p{0.8cm} p{2cm} Y c c c}
\toprule
\textbf{RQ} & \textbf{Themes} & \textbf{Categories} & \textbf{P1} & \textbf{P2} & \textbf{Total} \\
\midrule
    RQ1 & Introduction of 4DWW & Leadership support; Economic context; Pilot framing; Initial KPI monitoring; Experimentation & 29 & 4 & 33 \\
    
    & Adaptation to 4DWW & Coordination redesign; Communication restructuring; Meeting structure changes; Employee behavioral adaptation; Asynchronous collaboration practices & 79 & 12 & 91 \\

    & Institu-tionalization of 4DWW & Normalization; Personal life integration; Identity and employer branding; Stabilization of routines; Cohort effects on expectations; Change in expectations & 18 & 20 & 38 \\
\midrule
    RQ2 & Survival testing under pressure & Economic downturn; Market shifts; Customer pressures; Restructuring and reorganization; Ownership and shareholder pressure & 2 & 14 & 16 \\

    & Protective adaptation & Voluntary overtime; Longer working days; Friday work when needed; Temporary return to 5D schedules; Self-imposed effort to protect the model & 9 & 5 & 14 \\

    & Anticipated rollback consequences & Employee reactions and resistance; Turnover intentions; Organizational image & 0 & 14 & 14 \\
\bottomrule
\end{tabularx}
\end{table*}
\vspace{-4mm}
\subsection{Threats to Validity}
As with every qualitative case study, ours has limitations. In the following, we describe them according to the case study research guidelines by Runeson and Höst~\cite{Runeson.2009}.

\textbf{Construct Validity:} 
In this study, we investigated 4DWW introduction, adaptation, normalization, survival testing, and survival. A potential construct validity threat concerns the operationalization of these abstract constructs. While the first three constructs were informed by prior literature on organizational change and the 4DWW, the concepts of survival testing and survival emerged inductively from the data. We define survival testing as periods surviving which an established 4DWW is challenged by changing organizational or external conditions. In contrast, 4DWW survival refers to the organization’s continued maintenance of the arrangement despite such pressures.
To strengthen construct validity, we relied on multiple sources of evidence, including two rounds of semi-structured interviews conducted four years apart, and discussions among the authors to refine code definitions and interpretations. The longitudinal design also enabled us to distinguish between adaptations associated with the initial transition to the 4DWW and those related to its continued maintenance under changing conditions.

\textbf{Internal Validity:} In the first interview phase, a single researcher coded all transcripts, which introduces single-code bias. However, the conducted LLM consistency check provides a quality assured measure. For the second phase, we reduced bias risks by applying double-coding procedure and, again, LLM based consistency checks. The observation window of our study spans concurrent changes, including market and AI pressure or re-organizational measures. Thus, we report co-occuring conditions rather than attributing the observed shifts causally to the 4DWW. Noteworthy, we relied on interviewees' memory in recollection of the events. The validity of the results is addressed by interviewing multiple people. 

\textbf{External Validity:} This study covers a single organization and the results are based on 15 interviews, which obviously limits generalization. Following Yin~\cite{Yin.2018}, we claim analytical rather than statistical generalization: our case context and the created artifacts, especially the 4DWW survival matrix, support transfer toward similar contexts, especially agile oriented organizations.  

\textbf{Reliability:} We made the research protocol, including the codebook, interview guides, and detailed information about the LLM annotator approach available to support replication and auditability. We are aware that our approach of a two-interview overlap and the LLM annotator consistency check is a too small sample for calculating a stable estimate for inter-coder agreement coefficient. The cross-phase interview consistency relies on a shared instrument core and the same analysis process applied. 

\section{Results}
\label{sec:Results}
In this Section, we answer our research questions. First, we cover \textbf{RQ 1: How is a four-day workweek introduced, adapted, and institutionalized in an agile software organization?}. Here, we clustered the results among the introduction of the 4DWW, individual adaptions to the 4DWW, and the institutionalization of the 4DWW. For \textbf{RQ2: How does a four-day workweek survive organizational and external pressures, and what factors may lead to its rollback?}, we identified three different cluster. We start with the results related to the survival testing of the 4DWW under pressure conditions and information about protective adaptions, before we present anticipated effects of a rollback to a 5DWW. 

\subsection{Establishing the Four-Day Workweek}
\subsubsection{
Introduction of the Four-Day Workweek.}
The 4DWW was introduced at Pied Piper in 2021, after the Covid-19 pandemic had moved the software teams to fully remote work in 2020. Interviewees consistently located the introduction in a period of favorable business conditions. Market pressure was low, the company performed well, and the pandemic had benefited rather than strained the business. Looking back from 2026, ID15 described the starting point as oriented toward corporate values: \textit{``At introduction it was strongly values-driven. [...] The market pressure simply was not there. On the contrary, through the pandemic we had clear economic success; we were one of the beneficiaries in the market''.} Participants tied the decision to the organization's confidence in its own performance data. As ID9 explained, \textit{``I imagine that step would never have been taken if we hadn't assumed that we have relatively good KPIs and know how well we function.''} The stated motivation was cultural rather than operational: The 4DWW was meant to shape an organizational identity attractive to current and future employees. 

The introduction was gradual rather than a single cut. The teams first moved to fully remote work, then received a free Friday afternoon, and only later a full free day of own free choice. The reduced-hour arrangement was declared a test phase for one to two years. ID2 came back to this time-frame: \textit{``It started as: you are all at home, and we are giving you half of Friday off. That was the first step. [...] The next step was the test phase of the four-day week, declared for one or two years. [...] It was very fluid in that special time.''} The reduced-Friday step acted as an unplanned pilot that made the full transition thinkable; without it, ID9 argued, the organization \textit{``would never have known how effective we actually can be.''} Several participants read the trial framing as deliberate risk management. The pilot label persisted longer than first announced, which some interpreted as the organization keeping a path back to a 5DWW (ID3). 

The 4DWW introduction was managed as a monitored change process. Interviewees described continuous measurement against key performance indicators tracking deliveries and service quality through biweekly pulse surveys, and cross-site task force that tracked how the operating working model performed. ID3 named the metrics directly: \textit{``It is mainly about checking whether we still answer the same number of tickets, where we are faster, where we are slower, customer satisfaction, and so on.''} ID6 framed the whole effort in change-management terms: \textit{``It is essentially a change-management process. You need good communication, which was there, and checking along the whole way that you are on the right track, which was communicated and measured well.''} Monitoring extended to agile iteration planning activities itself. To keep work manageable while team members were absent for a day, one team temporarily switched from two-week to one-week iterations (ID2). The introduction also rested on a broad commitment to making the model work. As (ID6) mentioned: \textit{``Everyone had in the back of their mind: let us do everything to make it work. Even the attitude, I work half an hour longer today, but in return I have my Friday off.''}
\vspace{-5mm}
\subsubsection{Behavioral, Collaborative, and Attitudinal Adaptations to the Four-Day Workweek.}
Operating on four days required the teams to redesign how they coordinate, communicate, and run meetings, and it changed how individuals worked. Across roles and across both interview phases, interviewees named coordination as the central challenge. The reduced week coupled with remote work removed the co-located, ad-hoc coordination that the office had supplied. As ID3 described it, \textit{``Very often it used to be: we are all in the office anyway, so let's just disappear into the meeting room for an hour and discuss it together. Or I quickly come over and interrupt you because I need something. Those options are suddenly gone. You realize that if I send a chat message, the person will answer it too. Or instead of going to the meeting room spontaneously, I check when the other person is available.''} Teams compensated with tighter planning, explicit handover, and coverage rules so that no function was unavailable for a full day.

The reduced working time under the 4DWW affected teams differently. Knowledge-work teams gained efficiency by protecting focus time, whereas high-volume task-processing functions such as Finance needed two to three months before the new model became workable. The constraint also prompted teams to reassess meetings, questioning which ones added value and how they should be time-boxed (ID15).

Communication shifted from synchronous to written and asynchronous. What participants described in the first interview phase as a precondition for the 4DWW they later named as a deliberate practice: writing things down, sharing in advance, and asking whether a meeting is warranted at all (ID13). Meeting load was concentrated on the core days. ID2 reported that, because colleagues took Friday and Monday off, larger meetings had to be scheduled mid-week, producing \textit{``meeting marathons''} from Tuesday to Thursday. By 2026 teams had cut meetings to a minimum and reserved co-working days for focused work. Synchronous engineering practices were affected directly. As ID11 noted, \textit{``There is less time for collaboration and more async collaboration. People in our business usually do pair programming, but this is now not possible all the time. We still want parallel work streams, and then you only have one person maybe available.''} 

At the individual level, participants described the 4DWW as conditional on working more efficiently. Effectiveness and automation were framed as the explicit price of keeping the 4DWW. As ID9 explained, \textit{``It was made clear from the start that people still deliver and that we do not need a total change of our goals, but a move toward more efficiency. Automation goals had started earlier; with the four-day week it was clear to everyone: we really have to watch effectiveness.''} Participants connected this efficiency focus to a perceived stability of team output, which they treated as the precondition for keeping the 4DWW.
\vspace{-5mm}
\subsubsection{Institutionalization of the Four-Day Workweek.}
Over time the 4DWW receded as an active topic and became the way work is organized. Participants described trust and self-organization as the cultural default, with the arrangement no longer requiring justification in everyday work. The 4DWW was also written into how the company hires: one interview stage for candidates is dedicated to company values and mindset (ID11). ID15 framed the institutionalized arrangement in cultural terms: \textit{``The [4DWW] paid in strongly on care and on trust. Care in the sense that, when it was introduced, a lot of attention went to how it is implemented: what conditions employees need at home for it to work, and that they do not wear themselves out in the four days.''} 

Employees built their personal lives around the additional day off. In the first interview phase, the interviewees framed the day-off as a recovery and health benefit. As ID7 explained, \textit{``One thing we did not talk about, and I believe still has an effect, is health and stress. I feel I have more recovery when needed, with one day more a week where I am off anyway, compared to a full five-day week. I rarely call in sick.''} Four years later the same logic appears as constitutive part of the employment relationship. ID13 tied the day directly to a long-term commitment: \textit{``I am doing a Master's degree. [...] For me the four-day week has a major advantage: you can develop yourself and live out personal things more.''} This embedding made the off-day difficult to separate from the employment proposition itself. 

The 4DWW became a stable recruitment and retention asset. ID11 reported that the company had hired substantial new talent with the 4DWW in mind while holding output roughly constant, and that retention held in difficult periods. Participants attributed part of the company's recruiting attractiveness to the 4DWW, while noting that it was hard to disentangle from the flexibility to work remotely (ID12). 

A growing share of staff had never had a five-day week at Pied Piper. As ID12 observed, \textit{``By now half of the colleagues started on a four-day week. In some cases we no longer have anyone who knows a different way of working from the past.''} For the transition cohort, the 4DWW had been a conditional privilege tied to meeting performance expectations; for new hires, it was simply the baseline. Participants noted that the latter colleagues were stricter with working hours and left precisely on the hour, unlike longer-tenured staff (ID14). Where staff carried a five-day rhythm into the 4DWW without the focus discipline described above, participants linked this to weaker team performance (ID11).

\subsection{Long-Term Survival of the Four-Day Workweek} 
\begin{figure}
    \centering
    \includegraphics[width=1\linewidth]{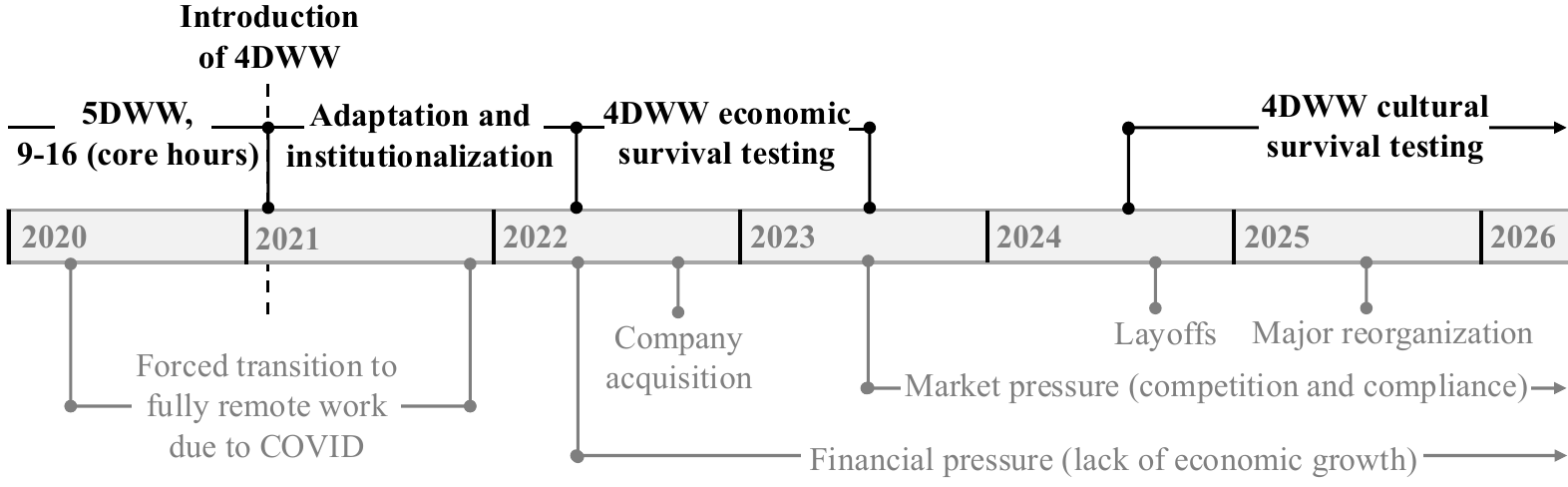}
    \caption{Events following the introduction of the 4DWW at the case company}
    \label{fig:timeline}
\end{figure}

In contrast to many studies focusing on transition to the 4DWW, we have revisited the case and interviewed the company representatives about the changes that happened during the last five years (see a summary in Figure~\ref{fig:timeline}). 

\vspace{-9mm}
\subsubsection{Survival testing under pressure conditions.}
All interviewees acknowledged that organizational context surrounding the 4DWW had changed considerably since its introduction in 2021. While the 4DWW was initially implemented during a period of economic growth and organizational optimism, the following years brought increasing uncertainty and pressure. Participants described a combination of economic downturns, market shifts, changing customer expectations, followed by ownership-related uncertainty, and organizational restructuring, that collectively challenged the long-term survival of the 4DWW arrangement.

Several interviewees linked these pressures to the company's acquisition and subsequent changes in ownership and shareholder expectations. These developments triggered multiple rounds of organizational restructuring and strategic realignment. Several interviewees (ID5, ID14) noted that one side of the implicit agreement underlying the 4DWW had changed: while employees had adapted to and embraced the new work model, new owners and business realities were not necessarily bound by the assumptions that existed when the arrangement was introduced. Under such conditions, the possibility of a rollback to a 5DWW appeared more plausible than at any point since its introduction. 
 
Interviewees also described growing market pressure driven by technological disruption and increased competition. Recent developments in AI and changes in the digital marketplace created additional innovation demands that competed with existing development priorities. As ID12 explained: \textit{``[Market changes] create innovation pressure. [...] There are always urgent topics appearing that disrupt planned sprint work. So, we constantly need quick coordination.''}

Customer-facing functions experienced these pressures particularly strongly. While development teams could often compensate through asynchronous work and flexibility, service-oriented teams faced greater challenges in maintaining responsiveness under reduced working hours. As ID13 argued, \textit{``I think [delayed customer support] is the strongest argument for why a five-day week could make sense in certain areas or why there could be pressure to return. Customer contact is crucial. We may have a platform business, but the human aspect still matters.''}

Importantly, interviewees emphasized that these pressures were not experienced uniformly across the organization. Operational constraints in different departments were not the same, resulting in varying levels of vulnerability and exposure to rollback pressures. 
\vspace{-5mm}
\subsubsection{Protective adaptation.} 
Rather than advocating for a return to a 5DWW, interviewees described a range of adaptive behaviors aimed at preserving 4DWW under increasingly demanding conditions. These included occasional longer working days, temporary returns to five-day schedules during particularly demanding periods, and greater personal responsibility for coordination and communication.

Notably, these adaptations were generally described as voluntary responses rather than management-imposed requirements. In this sense, employees actively absorbed part of the pressure generated by market demands, restructuring efforts, and customer expectations. As ID11 explained: \textit{``This 9-to-5 mentality is also not the best for this way of working. [...]. So, when you see a chat on a Friday, you're off, [...] but there is an open question and you are maybe the one who knows it, you [...] just drop a line and answer it.''} 

Several interviewees suggested that such behaviors reflected a strong collective commitment to preserving the 4DWW arrangement. In some cases, interviewees even expressed willingness to make personal sacrifices to retain the 4DWW. As ID13 reflected: \textit{``My very first thought would honestly be: how much money would I be willing to give up in order to keep a four-day week? I now know the value of having a four-day week. And it’s not even that I don’t want to work on the fifth day—I just want to do something else with that time.''}

\subsubsection{Anticipated Rollback Effects.}
Overall, interviewees anticipated that a rollback would negatively affect employee satisfaction, motivation, and organizational commitment, and damage the corporate image. As an interviewee (ID13) noted: \textit{``The four-day week helps retain employees. Without it, you lose that selling point''}. Many participants anticipated increased turnover intentions. Because 4DWW had become embedded in employees' personal routines, career decisions, and perceptions of organizational identity, a rollback was often viewed not simply as a scheduling change but as a fundamental alteration of the employment relationship and, in case of recent hires, even employment contracts.

We also learned that employees were concerned about the management expectations of increased output if returning to a 5DWW. Several participants noted that employees were already engaging in protective adaptation behaviors to sustain the 4DWW, with rather high workload. Reflecting on the considerable effort that employees invested in preserving the 4DWW, some interviewees anticipated that resistance to the rollback might be as strong as the commitment that initially supported the transition. An interviewee (ID15) suggested that employees might actively oppose a return to the five-day model. Another interviewee (ID12) exemplifies this: \textit{``The workload is already very high and cannot remain a permanent state. If we returned to five days, I would consciously slow myself down. I would start protecting my workload more actively.''} 

Interestingly, the interviewees rarely discussed 4DWW in isolation. It was consistently referred to as part of a broader package that combined reduced working time with temporal and locational flexibility. This is why, discussions about a potential rollback frequently prompted clarifying questions about whether this would also imply a return to fixed office attendance and traditional 9-to-5 schedules. Thus, the success of the current arrangement depends not only on the reduced number of working days but also on the ability to maintain flexible and individually optimized work routines. In this sense, reverting to a traditional five-day office-based workweek was perceived as substantially more disruptive than merely increasing the number of working days while retaining flexibility.

\section{Discussion}
\label{sec:Discussion}

\subsection{How Four-Day Workweek Evolves Beyond Implementation?}
In contrast to existing literature that largely treats the 4DWW as a decision and examines its short-term effects~\cite{Campbell.2023,Landwehr.2025}, with outcomes typically measured only months after the transition~\cite{Jahal.2024}, our longitudinal findings suggest that it is better understood as an evolving organizational arrangement that progresses through distinct phases. 

When Pied Piper decided to introduce the 4DWW, they piloted it first with different schedules, and, as a result, gained leadership support. This is consistent with studies that name managerial support and redesigned work practices as conditions for a successful transition towards a 4DWW~\cite{Munyon.2023}. The unplanned half-day pilot and continuous monitoring of product development outcomes lowered the perceived risk of the full transition and brought confidence in the company's own performance data to committed support.

After the introduction, employees went through the period of adaptation, focusing on work redesign and establishment of the necessary coordination mechanisms. Topp et al. \cite{Topp.2022} documented comparable coordination and remote-work changes shortly after the 4DWW introduction at Pied Piper, but only at a single point in time. Our longitudinal view shows these changes consolidating into deliberate practice, including asynchronous communication, explicit handover of tasks, and coverage rules. Removing an entire working day for an extended period reshaped synchronous agile practices more directly. For example, pair programming became intermittent, agile practices with longer time-boxes such as retrospectives concentrated mid-week. We also identified shortening the iteration length to one-week as an adaption of a team's agile approach in use. Participants in our study framed efficiency and automation as the explicit price of keeping the 4DWW, which is in line with the concentration gains and stress signals reported for compressed schedule in a software development context~\cite{Neumann.2025}.

After roughly a year after the introduction in 2021 and during the formal trial in 2022 and 2023, the 4DWW became institutionalized, business as usual. Pied Piper formed a new identity, while its employees formed new expectations. The institutionalization phase is the least examined in the literature on 4DWW. Prior work focuses on well-being and retention as the measured outcomes of the intervention (e.g., \cite{Landwehr.2025}), our study shows that outcomes can be turned into employment relationship, recruitment, and organizational identity, or, in other words, as constitutive expectations. 

In the following years, economic downturn and changing market conditions put new pressures on the company, forcing ownership change, and a call for efficiency optimization. Long-term viability of a 4DWW was challenged. Our study contributes with details about the ways Pied Piper navigated the period of survival testing, which is recognized as largely untested in 4DWW literature~\cite{Jahal.2024}. In particular, we found that contrary to the implicit assumption that increasing organizational pressure directly leads to a rollback to 5DWW, the Pied Piper story surfaces an internal protective mechanism. Our findings suggest that when an institutionalized 4DWW is threatened, employees and teams often absorb the pressure through adaptive behaviors intended to preserve the arrangement. 

\begin{figure}
    \centering
    \includegraphics[width=1\linewidth]{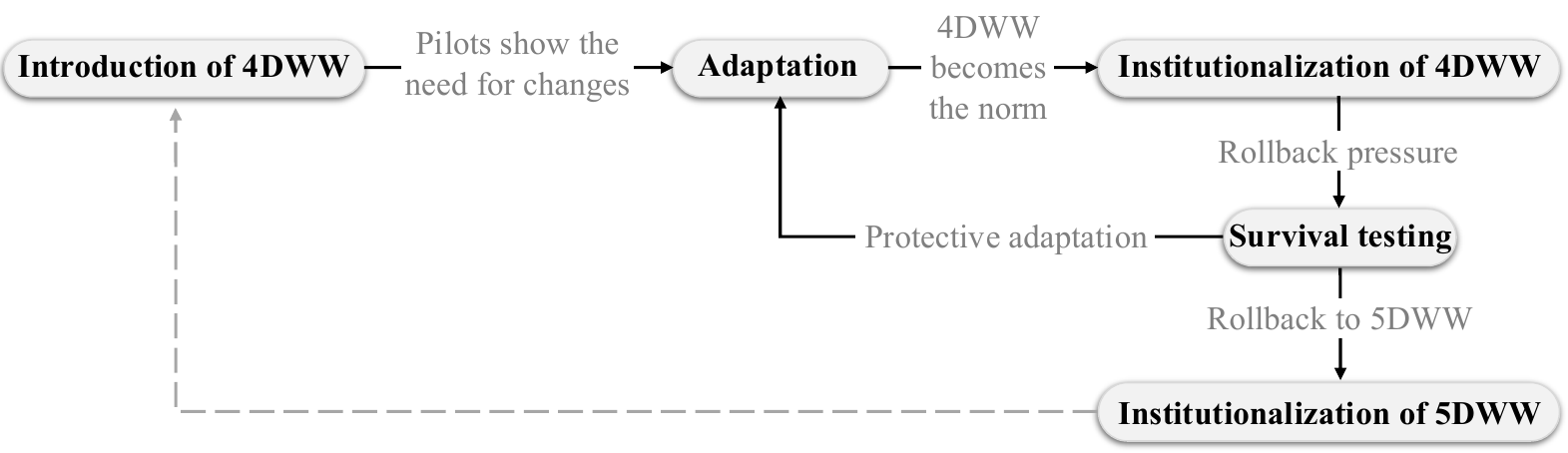}
    \caption{The Lifecycle of a Four-Day Workweek}
    \label{fig:lifecycle}
\end{figure}

In Figure~\ref{fig:lifecycle}, we summarize our results in a lifecycle of a 4DWW.

\vspace{-8mm}

\subsection{How Four-Day Workweek Survives?}
\begin{figure}
    \centering
    \includegraphics[width=1\linewidth]{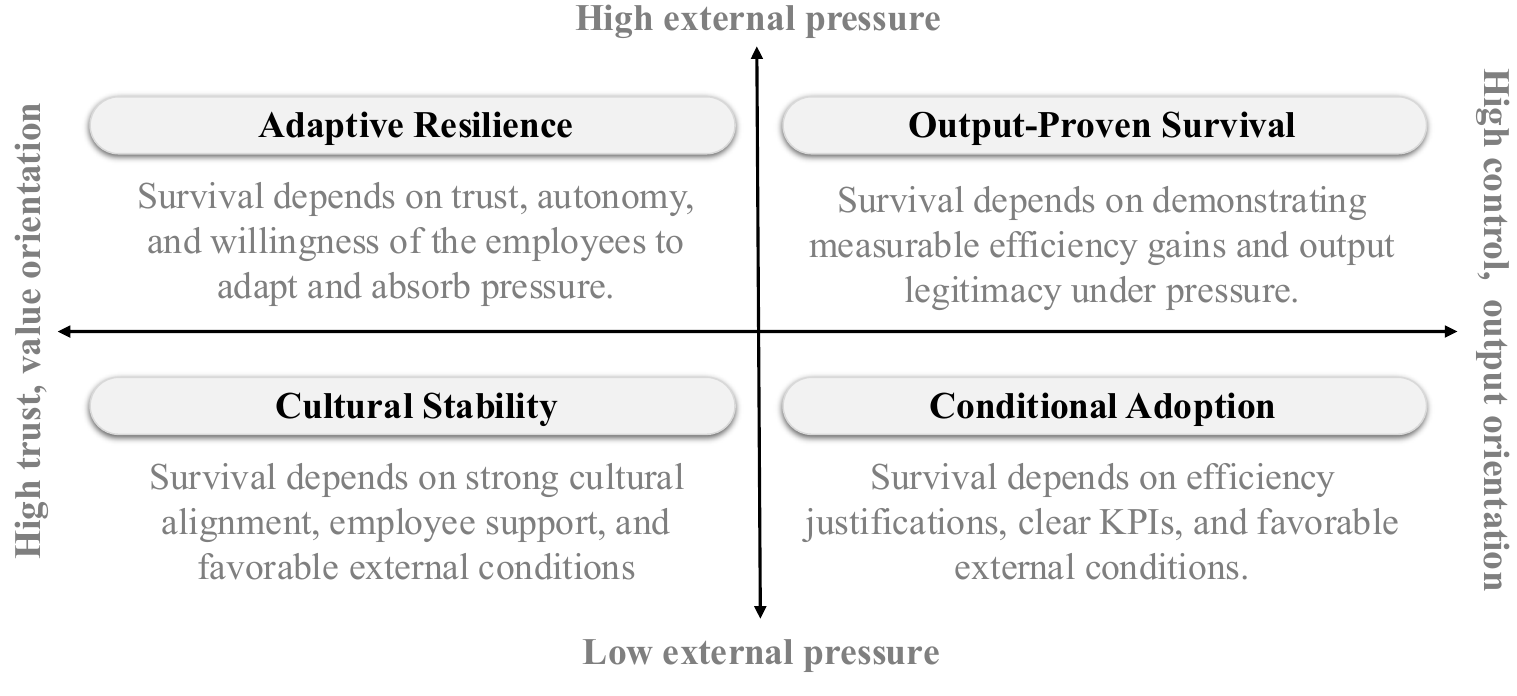}
    \caption{The Four-Day Workweek Survival Matrix}
    \label{fig:matrix}
\end{figure}

The lifecycle model presented in the previous section (See Figure~\ref{fig:lifecycle}) explains how a 4DWW evolves from introduction to normalization and eventually faces periods of survival testing. However, organizations might not respond to external pressures in a similar way. Even parts of the same organization might act differently. While some maintain the 4DWW with little resistance, others increasingly rely on performance justification, while yet others may eventually face rollback pressure. To explain these different trajectories, we propose the Four-Day Workweek survival matrix (see Figure~\ref{fig:matrix}). 

The matrix conceptualizes 4DWW survival as a function of two dimensions: 1) the level of external pressure faced by the organization and 2) the dominant rationale for introducing and later maintaining the 4DWW, ranging from trust- and value-oriented to control- and output-oriented management approaches. 

\vspace{-0.5mm}
\textbf{Conditional Adoption:} Organizations often introduce a 4DWW as an experiment whose continuation depends on measurable outcomes. In this quadrant, leadership evaluates the arrangement primarily through productivity indicators, efficiency gains, employee satisfaction measures, and business performance. The 4DWW arrangement survives as long as it can demonstrate value, but organizational commitment remains conditional rather than cultural. Thus, the arrangement is vulnerable if expected gains fail to materialize or if leadership priorities change.

\textbf{Cultural Stability:} In this quadrant, the 4DWW is aligned with organizational values and employee expectations. Leadership emphasizes trust, autonomy, well-being, and sustainable performance rather than short-term output maximization. The arrangement is supported by structural redesign of work practices rather than simple compression of working hours. However, long-term survival still depends on maintaining sufficient organizational slack and ensuring that productivity improvements are genuine rather than symbolic. The strongest threats to 4DWW survival here surface from external pressures. 

\textbf{Adaptive Resilience:} When value-oriented organizations face increasing market, customer, or organizational pressures, survival depends on the ability and willingness of employees to preserve the 4DWW. Rather than immediately abandoning the 4DWW, employees may engage in protective adaptation and absorb part of the pressure through voluntary overtime, temporary returns to five-day schedules, and intensified coordination. In this way, the 4DWW arrangement survives through trust, commitment, and collective adaptation rather than formal control mechanisms. 

\textbf{Output-Proven Survival:} When organizations face significant economic or market pressure while maintaining an output-oriented management philosophy, the legitimacy of the 4DWW depends on its ability to demonstrate measurable performance benefits. Leadership increasingly focuses on capacity utilization, productivity metrics, service responsiveness, and operational predictability. in this context, survival requires documented efficiency gains and evidence that reduced working hours support and not undermine organizational performance. However, rollback pressure becomes increasingly likely in this quadrant.

While we have based the 4DWW survival matrix on Pied Piper experiences to help explain why and how a 4DWW arrangement survives long-term, we believe it has practical value. The proposed matrix can be used as a diagnostic and strategic planning tool for organizations considering, implementing, or sustaining a 4DWW. It encourages leaders to assess both the external pressures that organizations are facing and the underlying rationale through which the 4DWW arrangement is legitimized. The matrix also provides an early warning mechanism for rollback risk. Organizations experiencing increasing pressures without either measurable performance legitimacy or strong adaptive commitment from their employees are more vulnerable to sustain the 4DWW.

\section{Conclusion \& Future Work}
\label{sec:Conclusion}
We studied how a 4DWW is introduced, adapted, institutionalized, and sustained in an agile software organization, drawing on two interview phases more than three years apart in the same company. The longitudinal
perspective shows that the 4DWW is better understood as an evolving arrangement than as a one-off intervention. After its introduction, the organization redesigned coordination and communication practices, the
arrangement became the institutional default, and it was then tested by ownership change, economic downturn, market and AI pressure, and organizational restructuring.  Our findings suggest that the survival of the 4DWW was not achieved solely through organizational policies, but also through individual and team-level efforts to protect the arrangement when its sustainability was challenged. We describe these efforts as protective adaptation and summarize the conditions that carry or threaten the model in the 4DWW survival matrix.

Future work should test the applicability of the matrix in other organizations, industries, and reduced-hour variants, as well as in other contexts, such as geographically and culturally distributed companies that our study did not cover. A second open question concerns the durability of protective adaptation: participants already reported high workload, which raises the question of whether voluntary effort can sustain a 4DWW over time or eventually erodes it.

\section*{Acknowledgments}
We would like to express our gratitude to Marvin Auf der Landwehr, who supported us in creating the interview guideline for Phase 1, conducted interviews, and participated in multiple discussions around the topic of this paper. Also, we would like to thank the participants from the case company for sharing their individual experiences and providing us valuable insights into their work and perspectives. 

%
%
%
 \bibliographystyle{splncs04}
 \bibliography{references}

\end{document}